\documentclass[useAMS]{mn2e}
\usepackage{times}
\usepackage{rotating}
\usepackage{graphicx}
\long\def\symbolfootnote[#1]#2{\begingroup%
\def\thefootnote{\fnsymbol{footnote}}\footnote[#1]{#2}\endgroup}
\newcommand{\gae}{\lower 2pt \hbox{$\, \buildrel {\scriptstyle >}\over {\scriptstyle
\sim}\,$}}
\newcommand{\lae}{\lower 2pt \hbox{$\, \buildrel {\scriptstyle <}\over {\scriptstyle
\sim}\,$}}

\begin{document}

\title[GRB magnetic fields]{Constraining the magnetic field in GRB
  relativistic collisionless shocks using radio data}  

\author[Barniol Duran]{R. Barniol Duran$^1$\thanks
{E-mail: rbarniol@phys.huji.ac.il}  \\
$^{1}$Racah Institute of Physics, The Hebrew University of Jerusalem, Jerusalem, 91904, Israel}

\date{Accepted; Received; in original form 2013 November}

\pubyear{2014}

\maketitle

\begin{abstract}

Using GRB radio afterglow observations, we calculate the 
fraction of shocked plasma energy in the magnetic field in relativistic
collisionless shocks ($\epsilon_B$). We obtained $\epsilon_B$ for 38 bursts
by assuming that the radio afterglow light curve originates in the external
forward shock and that its peak at a few to tens of days is due to the passage
of the minimum (injection) frequency through the radio band. 
This allows for the determination of the peak synchrotron flux of the external
forward shock, $f_p$, which is $f_p \propto \epsilon_B^{1/2}$.  The obtained value of
$\epsilon_B$ is conservatively a minimum if the time of the ``jet break'' is
unknown, since after the ``jet break'' $f_p$ is expected to decay with time faster
than before it. Claims of ``jet breaks'' have been made for a subsample of 23 bursts, for which
we can estimate a measurement of $\epsilon_B$. Our results depend on the blast wave
total energy, $E$, and the density of the circum-stellar medium (CSM), $n$, as
$\epsilon_B \propto E^{-2}n^{-1}$. However, by
assuming a CSM magnetic field ($\sim 10$ $\mu$G), we can express the lower
limits/measurements on $\epsilon_B$ as a density-independent ratio,
$B/B_{sc}$, of the magnetic field behind the shock to the CSM shock-compressed
magnetic field. We find that the distribution on both the lower limit on and
the measurement of $B/B_{sc}$ spans $\sim 3.5$ orders of magnitude and
both have a median of $B/B_{sc} \sim 30$.  This suggests that some
amplification, beyond simple shock-compression, is necessary to explain these radio
afterglow observations. 
\end{abstract}


\section{Introduction}

The external forward shock model (see, e.g., Rees \& M\'esz\'aros 1992; 
M\'esz\'aros \& Rees 1993, 1997; Paczy\'nski \& Rhoads 1993; Sari, Piran \& 
Narayan 1998) has been very useful for studying the afterglow emission of GRBs 
(Gamma-ray bursts; for a review, see, e.g., Piran 2004).  
This model explains the long-lasting emission in X-ray, optical and radio
bands as synchrotron emission (e.g., Sari et al. 1998, Granot \& Sari 2002) 
from circum-stellar medium (CSM) electrons that have been accelerated to high Lorentz Factors
(LFs) by the shock produced in the interaction between the GRB jet and the
CSM (for synchrotron-self-Compton emission, see, e.g., Sari \& Esin 2001, Nakar, Ando \& Sari
2009).  One of the main questions that remains unanswered is the origin of the 
magnetic field in the region where the electrons radiate, expressed as
$\epsilon_B$: the fraction of the total energy behind the shock in the magnetic 
field.  There are two ways to determine $\epsilon_B$: (1) Through afterglow 
modeling (e.g., Wijers \& Galama 1999; Panaitescu \& Kumar 2002; Yost et
al. 2003, Panaitescu 2005); and (2) through theoretical studies of magnetic field generation 
mechanisms (e.g., Medvedev \& Loeb 1999;  Milosavljevi\'c \& Nakar 2006; 
Sironi \& Goodman 2007; Goodman \& MacFadyen 2008; Couch, 
Milosavljevi\'c, Nakar 2008; Zhang et al. 2009; Mizuno et al.
2011; Inoue, Asano \& Ioka 2011).

The value of $\epsilon_B$ can be determined via afterglow modeling
for GRBs that have been {\it extensively} followed at different wavelengths. The 
reason is that the synchrotron spectrum consists of four power-law segments, 
which are divided by three characteristic frequencies, and observations should 
provide information about either the four different segments or the three 
characteristic frequencies (plus the peak flux) or a combination of the two
that allows us to determine the afterglow parameters {\it uniquely}.  In cases where extensive
afterglow follow-up at different wavelengths is not available, then various 
assumptions have to be made in order to determine the afterglow parameters;
these assumptions add to the uncertainty in the afterglow parameters.
In addition, past studies have made different assumptions, which have made
comparison between different works difficult.  The aim of this work is to
provide a systematic study, where we make the same assumptions for all GRBs in
our sample.  This allows us to obtain an $\epsilon_B$ distribution for which
one can draw meaningful conclusions.

In this paper, using radio afterglow observations, we calculate 
a lower limit on $\epsilon_B$. This value is obtained by assuming that the radio emission
is produced by the external forward shock and that the time when the radio
light curve starts to decay at a few
to tens of days is due to the passage of the minimum (injection) frequency through the radio 
band. The observed peak of the radio light curve indicates then the spectral
peak of the external forward shock, $f_p$, which depends on $\epsilon_B$ as
$f_p \propto \epsilon_B^{1/2}$ (e.g., Granot \& Sari 2002) and allows for an
estimate of $\epsilon_B$. The reason why $\epsilon_B$ is a minimum is because we conservatively
ignore if a ``jet break'' has been observed or not.  The ``jet break'' occurs when
the opening angle of the jet is approximately equal to the inverse of the
blast wave Lorentz factor and at this time the light curves show an achromatic break
(e.g., Rhoads 1999; Sari, Piran, Halpern 1999). After the ``jet break'', $f_p$
decays with time faster than before it, therefore, if the light curve peaks
after the ``jet break'', our  method would yield a larger value of $\epsilon_B$.

The identification of the ``jet break'' is not always a straightforward issue
(e.g., Liang et al. 2008; Kocevski \& Butler 2008; Racusin et al. 2009; Leventis et al. 2013).
This is the reason why we chose a conservative approach and estimated a lower
limit on $\epsilon_B$. Nevertheless, a subsample of GRBs in this work have
available ``jet break'' times, which have been estimated in combination with
optical and/or X-ray afterglows.  For this subsample, the lower limit can be
transformed into a measurement, but not without keeping in mind the
uncertainties mentioned above.  

The lower limits/measurements of $\epsilon_B$ in this study depend on the 
blast wave total energy and the density of the CSM.  However, assuming a 
magnetic field in the CSM, we can express our value of $\epsilon_B$ as a ratio of the magnetic
field behind the shock, $B$, to the CSM shock-compressed magnetic field,
$B_{sc}$, which turns out to be independent of CSM density.  
This ratio, $B/B_{sc}$, allows us to determine how much amplification -- 
beyond shock compression -- is needed. Since we are using mainly radio data, we make certain 
assumptions to determine $B/B_{sc}$ (as mentioned above). However, 
in this particular study, the only unknowns are the total energy in the blast
wave, which can be determined with some degree of confidence by knowing 
the observed gamma-ray radiated energy during the GRB, and the
seed (unshocked) magnetic field in the CSM, which is taken to be 
$\sim 10$ $\mu$G. The method presented in this paper provides a quick way to
estimate a lower limit/measurement of $\epsilon_B$ (and $B/B_{sc}$); 
it should not be used as a substitute for careful and
dedicated afterglow modeling, but it certainly provides a novel way to handle
a large number of GRBs easily. 

The present paper follows the work by Santana, Barniol Duran \& Kumar (2014), in which
they use optical and X-ray data to determine $\epsilon_B$ for a large number
of bursts.  The main idea of their work is to: 1. use the optical emission to
determine $\epsilon_B$ by assuming that this emission was produced in the
external forward shock, and 2. use the X-ray steep decay observed in many
X-ray light curves, which does {\it not} have an external shock origin, to place an
upper limit on $\epsilon_B$ by assuming that the external forward shock
emission is below the observed steep decay. The present work attempts to use
another wavelength, the radio band, to determine $\epsilon_B$ by yet another
method, and thus compare our findings with those of Santana et al. (2014) and
other authors. 

In Section \S \ref{Theory} we present how a measurement of $\epsilon_B$ (or a
lower limit) can be obtained using the peak of the radio light curve.  We
present our sample in \S \ref{Sample}, which is taken from Chandra \& Frail
(2012).  This sample is used to obtain a lower limit on $\epsilon_B$, and
also to obtain a measurement of $\epsilon_B$ for those bursts which have an
estimate of the jet break time. We present our results in the form of histograms
in \S \ref{Results}.  We discuss these results and present our conclusions in
\S \ref{final}.

\section{Finding $\epsilon_B$ through radio observations} \label{Theory}

In the external forward shock model (see, e.g., Sari et al. 1998) 
the injection (or minimum) frequency, $\nu_i$, which is the synchrotron frequency 
at which the injected electrons with the minimum LF radiate behind the 
shock, decreases with time as $\propto t^{-1.5}$.  At $\nu_i$, the specific
flux of the external shock is a maximum, which we denote as $f_p$.  At a few to ten days after
the burst, $\nu_i$ is predicted to be in the radio domain, $\nu_R$. Thus, radio
afterglow observations provide the best opportunity to determine 
the value of $f_p$.

\subsection{$\nu_i$ crosses the radio band before the jet break} \label{Theory_before_jet}

The radio afterglow light curve, for $\nu_R < \nu_i$, is predicted to rise slowly with time as
$\propto t^{1/2}$ (stay constant) for the case of constant CSM (wind profile), 
and to start decreasing at the time when  $\nu_i=\nu_R$, which we denote as
$t_i \equiv t(\nu_i=\nu_R)$ (referred to as $t_m$ in the notation of Sari et al. 1998).  
This is true if $t_i<t_j$, where
$t_j$ is the jet break time (Rhoads 1999; Sari, Piran, Halpern 1999).
The predicted synchrotron peak flux in the external
forward shock occurs at $\nu_i$ and it is given by (Granot \& Sari 2002)

\begin{equation} \label{peak_flux}
f_p = (9.93 \times 10^3 {\rm \mu Jy})(p+0.14)(1+z) \epsilon_B^{1/2} n_0^{1/2} E_{52} d_{L,28}^{-2},
\end{equation}
where $p$ is the power-law index for the energy distribution of injected
electrons, $E$ is the isotropic kinetic energy in the external shock, 
$n$ is the CSM density, $z$ is the redshift and $d_L$ is the luminosity
distance\footnote{The formula for $f_p$ in Leventis
et al. (2012), derived based on 1D relativistic hydrodynamic
simulations, is consistent with eq. (\ref{peak_flux}) within $\sim 30$ per cent. 
Given the level of uncertainty of our observables (see \S \ref{Sample}) and of
our assumptions, we will simply use eq. (\ref{peak_flux}).}.  
We use the convention $Q_x = Q/10^x$. Eq. (\ref{peak_flux}) is correct for a
constant density medium; however, we will express our results later in a
density-independent manner, and show that our method is independent of density 
stratification (see \S 4).   

If we can identify the time of the peak of the radio light curve, $t_p^{obs}$,
as $t_p^{obs}=t_i$, and the peak flux at this time is $f_p^{obs}$,
then $f_p=f_p^{obs}$ and this yields [see eq. (\ref{peak_flux})]

\begin{equation} \label{epsilon_B_peak_flux}
\epsilon_B = 
\left[\frac{f_p^{obs}}{9.93 \times 10^3 {\rm \mu Jy}} 
\frac{d_{L,28}^2}{(p+0.14)(1+z) E_{52}}\right]^2 \frac{1}{n_0}.
\end{equation}

\subsection{$\nu_i$ crosses the radio band after the jet break} \label{Theory_after_jet}

If $t_j < t_i$, then for $\nu_R < \nu_i$ the radio light curve
is predicted to rise as $\propto t^{1/2}$ (stay constant) for the case of constant CSM
(wind profile) and at $t_j$ to remain roughly
constant\footnote{After the jet break, for $\nu_R<\nu_i$, the radio light
curve has been shown analytically to decrease very slowly as $t^{-1/3}$ (Rhoads 1999; Sari et al. 1999), 
however, numerically, the light curve is found to be roughly constant (e.g.,
Zhang \& MacFadyen 2009).  In any case, we assume a roughly constant light
curve.} until $t_i$, when the light curve will
decrease as $\sim t^{-p}$ (Rhoads 1999; Sari et al. 1999).

We can find an equation analogous to eq. (\ref{epsilon_B_peak_flux}) for this case.  After the
jet break, the peak flux decreases as $f_p \propto t^{-1}$ independent of density medium (Sari et
al. 1999).  This same behavior of $f_p$ is found in detailed numerical
  simulations of a GRB jet interacting with a constant or a wind density medium
  (De Colle et al. 2012, van Eerten \& MacFadyen 2013).  At $t_i$, when the
light curve starts decreasing rapidly, the peak
flux is smaller than at $t_j$ and it is given by $f_p (t_i/t_j)^{-1}$.
If we can identify the time when the radio light curve
starts decreasing rapidly, $t_p^{obs}$, as $t_p^{obs}=t_i$, and
the flux at this time is $f_p^{obs}$, then this flux should be given 
by $f_p^{obs} = f_p (t_p^{obs}/t_j)^{-1}$.  This last expression yields, 
using eq. (\ref{peak_flux}), 

\begin{equation} \label{epsilon_B_peak_flux_2}
\epsilon_B = 
\left[\frac{f_p^{obs}}{9.93 \times 10^3 {\rm \mu Jy}} \frac{d_{L,28}^2}{(p+0.14)(1+z) E_{52}}\right]^2 \frac{1}{n_0}\left(\frac{t_p^{obs}}{t_j} \right)^2.
\end{equation}
To summarize, by identifying the time and flux at which the radio light curve
starts to decrease, $(t_p^{obs},f_p^{obs})$, and by determining the time of the jet break, 
we can use eq. (\ref{epsilon_B_peak_flux}) or eq. (\ref{epsilon_B_peak_flux_2}) 
to obtain a measurement of $\epsilon_B$. If the exact location of the jet
break is unknown, using eq. (\ref{epsilon_B_peak_flux}) would provide a
{\it lower limit} on $\epsilon_B$, which can be seen from eq. (\ref{epsilon_B_peak_flux_2}).

Before we continue, we would like to address the possibility that the peak of
the radio light curve is due to the passage of the self-absorption frequency, 
$\nu_a$ through the radio band, instead of the passage of $\nu_i$ as
considered above. Consider $\nu_R < \nu_a < \nu_i$.  In this case, $\nu_a$ is
a constant for a constant CSM, so we must consider a wind profile.  For this
profile, $\nu_a \propto t^{-3/5}$, and the passage of $\nu_a$ only makes the
light curve transition from $\propto t$ to $\propto t^0$; it does not make the 
light curve decrease. Therefore, a later passage of $\nu_i$ through $\nu_R$ 
(as considered above) is necessary to have a declining light curve. 
Alternatively, since $\nu_i$ decreases very rapidly, 
$\nu_R < \nu_a < \nu_i$ could transition quickly to 
$\nu_R < \nu_i < \nu_a$.  In that case, we would first get the passage 
of $\nu_i$ through $\nu_R$, which would make the light curve transition from 
$\propto t^{1/2}$ to $\propto t^{5/4}$ ($\propto t$ to $\propto
t^{7/4}$) for a constant CSM (wind profile), and then we would get the passage
of $\nu_a$ through $\nu_R$, which would make the light curve decrease. This 
last scenario is a definite possibility to explain the peak of radio
light curves; however, most radio light curves have a shallower rise, which 
strongly suggests the passage of $\nu_i$ through the radio band as the origin of their 
peak.

\section{Sample selection} \label{Sample}

We use the radio afterglow light curves of Chandra \& Frail (2012).  
They use a simple formula ($\propto t^{1/2}$ and $\propto t^{-1}$,
below and above the peak, respectively) to fit the radio light curves. They 
acknowledge that this fit may not be too accurate to represent the entire light curve,
however, it is good enough to determine the approximate values of the 
peak flux, $f_p^{obs}$, and the time of the peak, $t_p^{obs}$ (they use 
the notation $F_m$ and $t_m$, respectively, see their table 4).  These
  two values, instead of the precise temporal properties of the emission, are
  the crucial values needed in our analysis.

For the case when the radio light curve starts decaying {\it before} the jet break,
then it is easy to identify the fit done in Chandra \& Frail (2012) as an
indication of when $\nu_i$ crosses the radio band and use eq.
(\ref{epsilon_B_peak_flux}) to determine $\epsilon_B$ (see Section 
  \ref{Theory_before_jet}). This is true for
  the constant density medium case. For the wind case, the fit should have
  been $\propto t^0$ before the peak; however, since we are interested in the
  flux and time when the light curve starts to decrease, then this
  is also approximately valid for the wind case.
For the case when the radio light curve starts
decaying {\it after} the jet break, then the identification of $\nu_i$ is not as
straightforward.  For this case, strictly speaking, one should have allowed for
a different fit with three power-laws: $\propto t^{1/2}$, $\sim t^{0}$ and
$\propto t^{-p}$ for a constant density medium, or simply $\propto t^0$,
  $\propto t^{-p}$ for a wind medium, as described in the previous section.
However, the approximate fit done in Chandra \& Frail (2012) and their values of 
$t_p^{obs}$ and $f_p^{obs}$ will still approximately indicate the time when
$\nu_i$ crossed the radio band and the flux at that time.  Therefore, for this
case we can use eq. (\ref{epsilon_B_peak_flux_2}) to determine $\epsilon_B$. For
the case when $t_j$ is not known, we can use eq. (\ref{epsilon_B_peak_flux}) 
to determine a lower limit on $\epsilon_B$.

A detailed fit done to each of the radio light curves could in principle indicate the time of
the jet break and the time when $\nu_i$ crosses the radio band (and eliminate
a possible small source of error in our calculation, which arises from
differences in the fit done in Chandra \& Frail 2012 and the one described in
the previous section).  If more than
one radio band is available, then this task is even more certain.  However, in
cases when the radio data is sparse, or scintillation plays a major role (Frail
et al. 2000), then
one needs to rely on optical or X-ray data to extract the time of the jet
break.  In this particular study, we use the jet break values compiled in
the table 1 in Chandra \& Frail (2012), which are obtained using either
optical, X-ray or radio observations, and sometimes a combination of two or 
more bands.

We begin our sample selection with all 54 GRBs reported on table 4 of 
Chandra \& Frail (2012) for which there were enough data 
points to determine $t_p^{obs}$ and $f_p^{obs}$. Only 45 of these had a 
redshift\footnote{A redshift range $z=2-3.9$ is given for GRB 980329, we use $z=3$ as 
Chandra \& Frail (2012).}.  We calculate the bulk Lorentz factor of the blast
wave, $\Gamma$, at $t_p^{obs}$ for these 45 bursts\footnote{The blast wave LF at time $t_d$ (in days) is 
$\Gamma(t_d) = 8 (E_{52}/n_0)^{1/8} ((1+z)/2)^{3/8} t_d^{-3/8}$
(Blandford \& McKee 1976).}, and find that for 7 of them
$\Gamma(t_p^{obs}) < 2$.  For simplicity, we do not include them in our sample,
because a transition to the subrelativistic regime and/or a deviation 
from the jet geometry might be present in these 7 bursts. This yields a total
sample of 38 bursts\footnote{Some radio light curves have some data on more than one radio frequency 
and we use the frequency with the shortest peak time to ensure that the 
blast wave is still relativistic. In most cases this also corresponds to the 
largest $f_p^{obs}$, which provides a more conservative value of
$\epsilon_B$.}, for which we can obtain a {\it lower limit} on $\epsilon_B$ by using
eq. (\ref{epsilon_B_peak_flux}).

Of these 38 bursts, 19/38 have reported values of
$t_j$ and 7/38 have limits on $t_j$ (4 lower limits and 3 upper limits), see 
table 1 of Chandra \& Frail (2012). To determine a measurement of $\epsilon_B$, the 3 upper limits on $t_j$
are not useful; however, 3 out of the 4 lower limits on $t_j$ which are larger than
$t_p^{obs}$ are useful, since they indicate $t_p^{obs} < t_j$ and
eq. (\ref{epsilon_B_peak_flux}) can be used.  This yields a total sample of 23
GRBs, for which we can obtain a {\it measurement} of $\epsilon_B$ with 
eqs. (\ref{epsilon_B_peak_flux}) and (\ref{epsilon_B_peak_flux_2}) 
for $t_p^{obs} < t_j$ and $t_p^{obs} > t_j$, respectively. 

\begin{figure}
\begin{center}
\includegraphics[width=8cm, angle = 0]{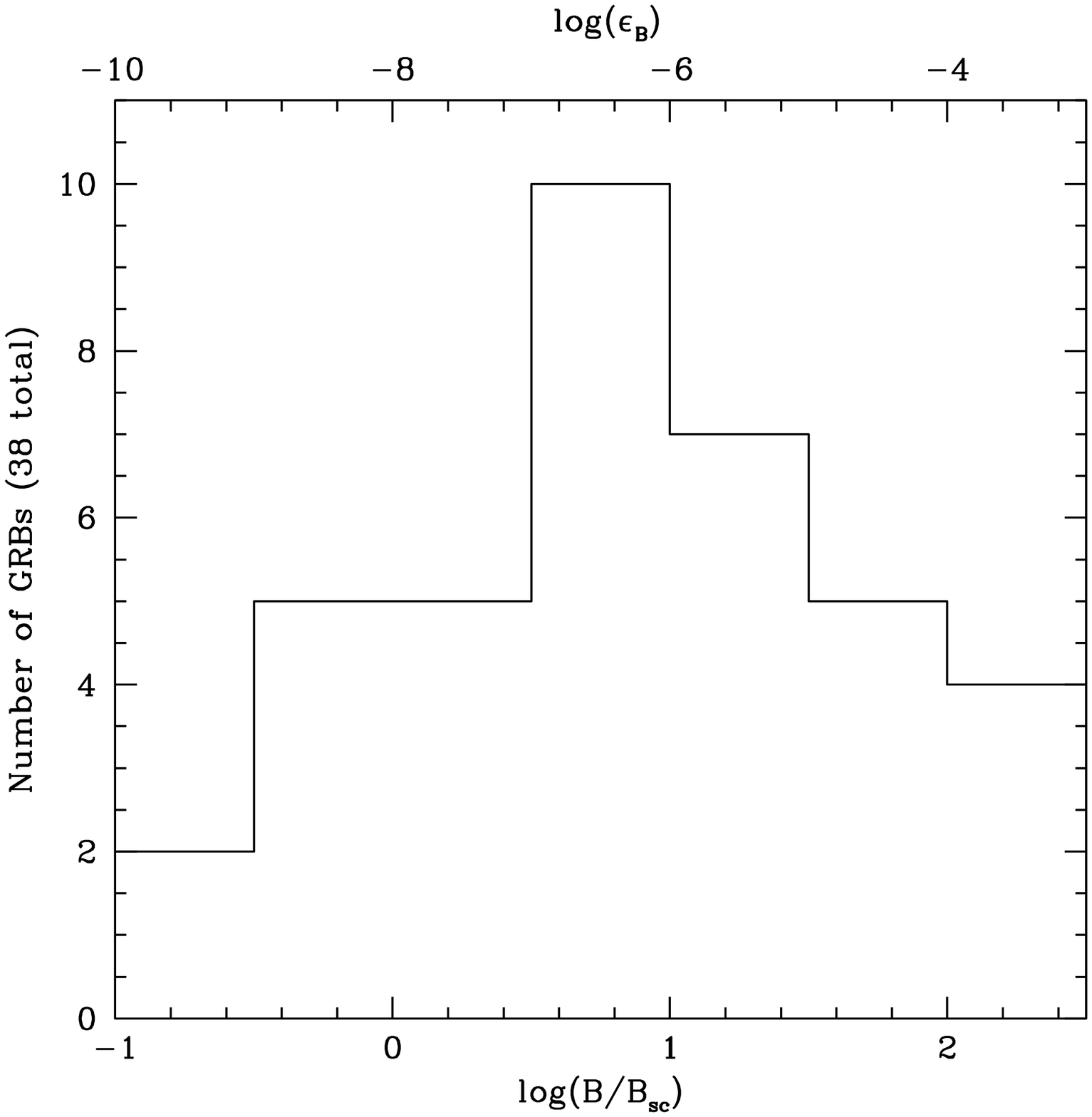}
\includegraphics[width=8cm, angle = 0]{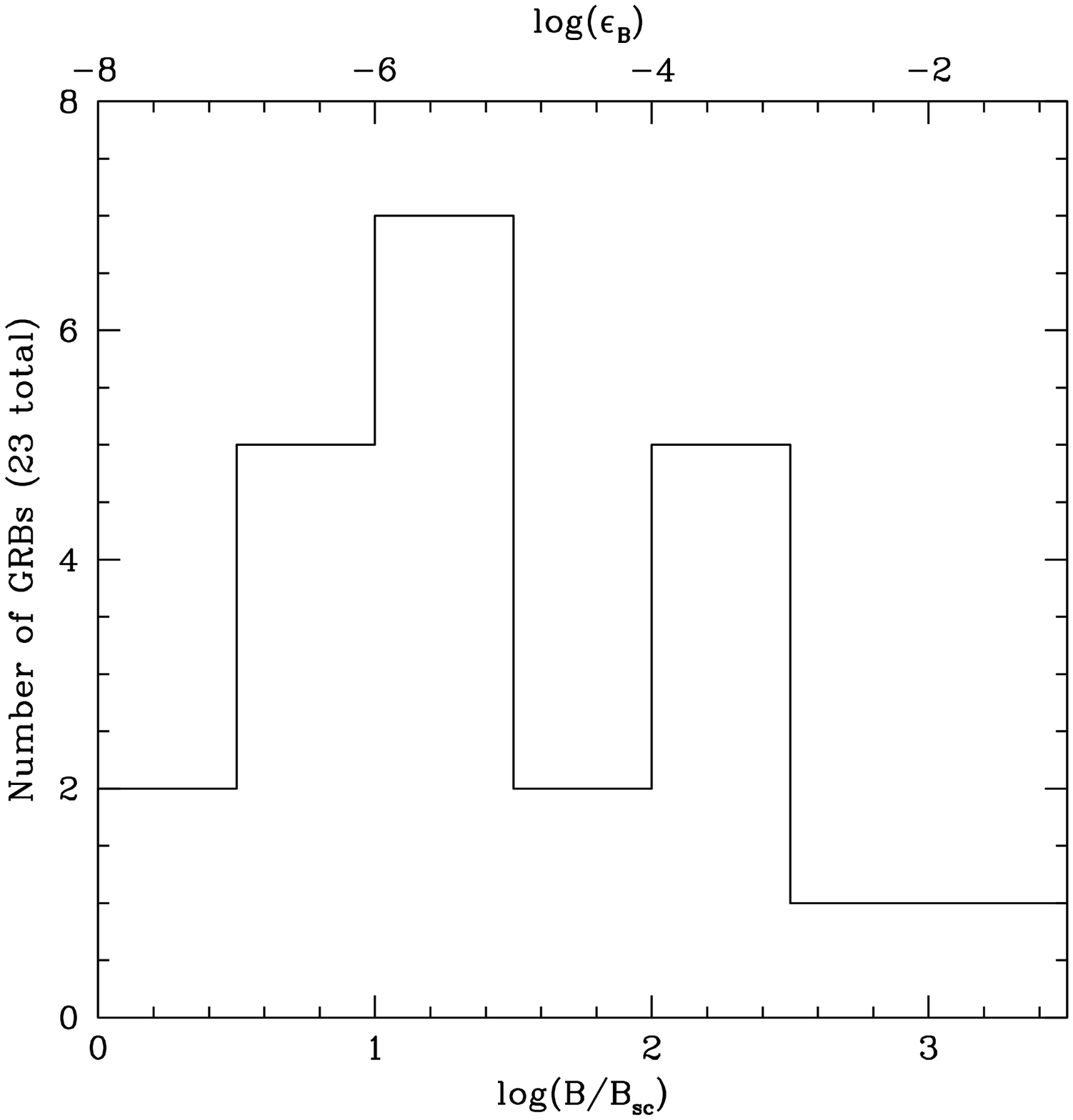}
\end{center}
\caption{The observed radio light curve peak flux indicates 
the time when the injection frequency crosses the observed radio band, 
and yields a lower limit on $\epsilon_B$ (upper panel).  It yields a 
measurement of $\epsilon_B$ (lower panel) for those bursts for which a ``jet
break'' has been identified. We present the histogram of lower limits of 
$\epsilon_B$ (see upper-horizontal axes) obtained 
using eq. (\ref{epsilon_B_peak_flux}) for the sample of GRBs in Table
\ref{table1} (upper panel), and the histogram of the measurements of 
$\epsilon_B$ obtained using eqs. (\ref{epsilon_B_peak_flux}) 
and (\ref{epsilon_B_peak_flux_2}), depending if $t_p^{obs} < t_j$ or
$t_p^{obs} > t_j$, for the sample of GRBs in Table 
\ref{table2} (lower panel). We have used $p= 2.4$, $E=5 E_{\gamma}^{iso}$ 
and $n=1$ cm$^{-3}$.  We also express $\epsilon_B$ as $B/B_{sc}$ (independent 
of CSM density), which is the ratio of the magnetic field behind the shock
to the expected shock-compressed magnetic field, assuming an unshocked
CSM field of $B_0=10$ $\mu$G (see lower-horizontal axes). We note that 
$B/B_{sc}$ does depend on the assumed value of $B_0$ as 
$B/B_{sc} \propto (B_0/10\mu$G$)^{-1}$.  This choice of $B_0$ yields $B/B_{sc}<1$ 
in the upper panel, which is not physically possible, but it is consistent, 
since we are simply reporting a lower limit for a choice of $B_0$, and 
$B_0=1\mu$G would yield $B/B_{sc}>1$ for all bursts.}
\label{fig1} 
\end{figure}

We use the observed peak flux $f_p^{obs}$ at $t_p^{obs}$ and the value of
$t_j$ from Chandra \& Frail (2012). We treat both
samples, the one with 38 GRBs (lower limits on $\epsilon_B$) 
and its subsample of 23 GRBs with $t_j$ (measurements of $\epsilon_B$),
separately. We take the isotropic kinetic energy to be 
$E = 5 E_{\gamma}^{iso}$, where $E_{\gamma}^{iso}$ is the isotropic gamma-ray
radiated energy during the prompt emission, so that the efficiency in
producing gamma-rays is $\sim 20$ per cent. We also take
$p = 2.4$ (Curran et al. 2010), however, its exact value does not affect
our results. Finally, the density is a free parameter, and we take $n=1$ cm$^{-3}$ to
display our results, which means that the histograms and tables can be viewed as
displaying the quantity $\epsilon_B n$. We report the data of each GRB of our sample 
of 38 bursts (23 bursts) in Table \ref{table1} (Table \ref{table2}) and we 
present a histogram of the lower limits (measurements) of $\epsilon_B$ in Figure \ref{fig1}.


\begin{table*}
\begin{center}
\begin{tabular}{cccccccc}
\hline
GRB	&	$z$	&    $d_{L,28}$ &	$E_{\gamma,52}^{iso}$ & $f_p^{obs}$	&	$t_p^{obs}$
& $\log(\epsilon_B)$	& $B/B_{sc}$  \\
\hline
021004	&2.33	&	5.9	&	3.80	&	1308	&	4.1	&	-3.1	&272\\
970508	&0.84	&	1.6	&	0.71	&	958	&	37.2	&	-3.7	&142\\
090313	&3.38	&	9.2	&	4.57	&	435	&	9.4	&	-3.7	&139\\
030329	&0.17	&	0.25	&	1.80	&	59318	&	5.8	&	-3.7	&133\\
000301C	&2.03	&	4.95	&	4.37	&	520	&	14.1	&	-4.2	&73\\
981226	&1.11	&	2.3	&	0.59	&	137	&	8.2	&	-4.7	&44\\
080603A	&1.69	&	3.94	&	2.20	&	207	&	5.2	&	-4.7	&41\\
091020	&1.71	&	4	&	4.56	&	399	&	10.9	&	-4.8	&39\\
100814A	&1.44	&	3.2	&	5.97	&	613	&	10.4	&	-4.9	&33\\
000418	&1.12	&	2.36	&	7.51	&	1085	&    18.1	&	-5.0	&29\\
090423	&8.26	&	26.36	&	11.00	&	50	&	33.1	&	-5.1	&26\\
980703	&0.97	&	1.97	&	6.90	&	1055	&	9.1	&	-5.2	&23\\
071010B	&0.95	&	1.92	&	2.60	&	341	&	4.2	&	-5.4	&19\\
011211	&2.14	&	5.28	&	6.30	&	162	&	13.2	&	-5.5	&17\\
000926	&2.04	&	4.98	&	27.00	&	629	&	12.1	&	-5.6	&14\\
991208	&0.71	&	1.34	&	11.00	&	1804	&7.8&-5.7&13\\ 
090715B	&3	&	7.98	&	23.60	&	191	&	9.2	&	-6.0	&9.7\\
020819B	&0.41	&	0.69	&	0.79	&	291	&	12.2	&	-6.0	&9.3	\\
030226	&1.99	&	4.8	&	12.00	&	171	&	6.7	&	-6.1	&8.3\\
050603	&2.8	&	7.3	&	50.00	&	377	&	14.1	&	-6.2	&8.0\\
071003	&1.6	&	3.7	&	32.40	&	616	&	6.5	&	-6.2	&7.6\\
050820A	&2.62	&	6.8	&	20.00	&	150	&	9.4	&	-6.2	&7.2\\
070125	&1.55	&	3.54	&	95.50	&	1778	&	13.6	&	-6.3	&6.9\\
090328	&0.74	&	1.41	&	10.00	&	686	&	16.1	&	-6.4	&5.9	\\
990510	&1.62	&	3.7	&	18.00	&	255	&	4.2	&	-6.5	&5.6\\
010921	&0.45	&	0.77	&	0.90	&	161	&	31.5	&	-6.5	&5.5	\\
011121	&0.36	&	0.59	&	4.55	&	655	&	8.1	&	-7.1	&2.8	\\
090424	&0.54	&	1	&	4.47	&	236	&	5.2	&	-7.1	&2.6\\
050904	&6.29	&	19.23	&	130.00	&	76	&	35.3	&	-7.3	&2.2\\
100414A	&1.37	&	3.03	&	77.90	&	524	&	8	&	-7.4	&2.0\\
980329	&3	&	7.97	&	210.00	&	332	&	33.5	&	-7.4	&1.9\\
090323	&3.57	&	9.84	&	410.00	&	243	&	15.6	&	-8.0	&0.9\\
020405	&0.69	&	1.29	&	11.00	&	113	&	18.2&	-8.2	&0.8\\
970828	&0.96	&	1.9	&	29.60	&	144	&	7.8&	-8.3	&0.7\\
000911	&1.06	&	2.2	&	88.00	&	263	&	3.1	&	-8.5	&0.5	\\
051022	&0.81	&	1.6	&	63.00	&	268	&	5.2	&	-8.6	&0.5	\\
010222	&1.48	&	3.34	&	133.00	&	93	&	16.8	&	-9.2	&0.2	\\
090902B	&1.88	&	4.5	&	310.00	&	84	&	14.1	&	-9.6	&0.1\\
\hline
\end{tabular}
\end{center}
\caption{The condition that the observed radio light curve peak flux, $f_p^{obs}$ 
(in $\mu$Jy, from Chandra \& Frail 2012),  should be smaller than or equal to the peak synchrotron flux
gives a lower limit on $\epsilon_B$ [eq. (\ref{epsilon_B_peak_flux})], 
assuming $n=1$ cm$^{-3}$.  $\epsilon_B$ is also 
written as a ratio of the magnetic field behind the shock to the expected
shock-compressed magnetic field, $B/B_{sc}$, which is independent of CSM
density.  We assumed that the unshocked CSM field is 10 $\mu$G.  
The redshift is $z$, the luminosity distance (in units of 
$10^{28}$ cm) is $d_{L,28}$, the k-corrected observed isotropic gamma-ray energy (in units
of $10^{52}$ erg) is $E_{\gamma,52}^{obs}$ and the observed time of the peak (in days) is
$t_p^{obs}$ (Chandra \& Frail 2012). All observed radio frequencies are at
8.46 GHz, except for: GRB 021004 (15 GHz), GRB 030329 (43 GHz), 
GRB 100814A (7.9 GHz), GRB 980703 and GRB 010921 (4.86 GHz), GRB 070125 (22.5
GHz), and GRB 011121 (8.7 GHz).}
\label{table1}
\end{table*}

\begin{table*}
\begin{center}
\begin{tabular}{ccccccccc}
\hline
GRB & $t_p^{obs}$ & $t_j$ & $\log(\epsilon_B)$	& $B/B_{sc}$ & SFR & $\log(M_{gal})$ & SSFR & Morphology \\
\hline
090313	&	9.4	&	0.79	&	-1.5	&	1655	&		&		&		&		\\
090715B	&	9.2	&	0.21	&	-2.7	&	427	&		&		&		&		\\
{\it 021004}&	4.1	&       7.6	&	-3.1	&	272	&	29	&	10.2	&	1.83	&	S	\\
970508	&	37.2	&	25	&	-3.3	&	212	&	1.14	&	8.52	&	3.44	&	S	\\
000301C	&	14.1	&	5.5	&	-3.4	&	186	&		&		&		&		\\
{\it 030329}&	5.8	&       9.8	&	-3.7	&	133	&	0.11	&	7.74	&	2	&		\\
011211	&	13.2	&	1.77	&	-3.7	&	129	&	4.9	&	9.77	&	0.83	&	M	\\
000926	&	12.1	&	1.8	&	-4.0	&	97	&	2.28	&	9.52	&	0.69	&	M	\\
030226	&	6.7	&	0.84	&	-4.3	&	66	&		&		&		&		\\
{\it 000418}&	18.1	&       25	&	-5.0	&	29	&	10.35	&	9.26	&	5.69	&		\\
980703	&	9.1	&	7.5	&	-5.1	&	28	&	16.57	&	9.33	&	7.75	&		\\
{\it 090423}&	33.1	&      $>$45	&	-5.1	&	26	&		&		&		&		\\
070125	&	13.6	&	3.8	&	-5.2	&	25	&		&		&		&		\\
990510	&	4.2	&	1.2	&	-5.4	&	20	&		&		&		&		\\
011121	&	8.1	&	1.3	&	-5.5	&	17	&	2.24	&	9.81	&	0.35	&	D	\\
090328	&	16.1	&	9	&	-5.9	&	11	&	3.6	&	9.82	&	0.54	&		\\
050820A	&	9.4	&	7.35	&	-6.0	&	9	&		&		&		&		\\
090424	&	5.2	&	1.6	&	-6.1	&	8	&		&		&		&		\\
020405	&	18.2	&	1.67	&	-6.1	&	8	&	3.74	&	9.75	&	0.67	&	A, M	\\
{\it 010921}&	31.5	&  	33	&	-6.5	&	6	&	2.5	&	9.69	&	0.51	&	D	\\
010222	&	16.8	&	0.93	&	-6.7	&	4	&	0.34	&	8.82	&	0.51	&	S	\\
970828	&	7.8	&	2.2	&	-7.2	&	2	&	0.87	&	9.19	&	0.56	&	A, M	\\
{\it 090323}&	15.6	&      $>$20	&	-8.0	&	1	&		&		&		&		\\
\hline
\end{tabular}
\end{center}
\caption{The observed radio light curve peak flux, $f_p^{obs}$ (in $\mu$Jy) is
used to obtain a measurement of $\epsilon_B$ using 
eqs. (\ref{epsilon_B_peak_flux}) and (\ref{epsilon_B_peak_flux_2}), 
depending if $t_p^{obs} < t_j$ or $t_p^{obs} > t_j$. This can be done for 23 GRBs 
out of our sample of 38 (see Table
\ref{table1}) for which a value of $t_j$ (in days) has been reported in
Chandra \& Frail (2012).  We assume $n=1$ cm$^{-3}$. GRBs for which $t_p^{obs} < t_j$ are
displayed in italics. $\epsilon_B$ is also 
written as a ratio of the magnetic field behind the shock to the expected
shock-compressed magnetic field, $B/B_{sc}$, which is independent of CSM
density.  We assumed that the unshocked CSM field is 10 $\mu$G.  
The GRB host galaxy star formation rate is SFR, in $M_{sun}$ per year; 
its mass is $M_{gal}$, in units of $M_{sun}$; its specific star formation
rate is SSFR$=$SFR$/M_{gal}$, in units of Gyr$^{-1}$ (Savaglio et al. 2009,
see footnote 6) and its morphology is given in the
last column (S=Spheroid, M=Merger, D=Disk, A=Asymmetric) from the sample
presented in Wainwright, Berger \& Penprase (2007) using the notation in 
Savaglio et al. (2009).}
\label{table2}
\end{table*}

\section{Results} \label{Results}

In order to eliminate the uncertainty on density, which 
might vary over many orders of magnitude (Soderberg et al. 2006), we can also
write $\epsilon_B$ as a ratio of the magnetic field
behind the shock, $B$, to the expected shock-compressed CSM field, $B_{sc}$.

Since $\epsilon_B = (B^2/8 \pi)/(4 n m_p c^2 \Gamma^2)$, 
where $m_p$ is the proton mass and $c$ is the speed of light, 
and $B_{sc} = 4 \Gamma B_0$, where $B_0$ is the
unshocked CSM field, which we assume to be $B_0 \sim 10$ $\mu$G, we
find $\epsilon_{B,sc} = B_0^2/(2 \pi m_p c^2 n) \sim 10^{-8} n_0^{-1}$.
With this, we can translate $\epsilon_B$ to $B/B_{sc}$ as 
$B/B_{sc}=(\epsilon_B/\epsilon_{B,sc})^{1/2}$, 
where $\epsilon_B$ is the lower limit (measurement) obtained 
for our 38 bursts sample (23 bursts sample) and reported in 
Table \ref{table1} (Table \ref{table2}). 
This ratio is independent of the type of CSM medium, whether it is a constant
or wind CSM. However, it does depend on the assumed value of $B_0$ as 
$B/B_{sc} \propto (B_0/10\mu$G$)^{-1}$. With our definition of $B/B_{sc}$, it is possible
that our very small lower limits on $\epsilon_B$ yield $B/B_{sc}<1$.  This is
not physically possible, but it is consistent, since we are simply reporting a
lower limit for a choice of $B_0$. Choosing $B_0=1$ $\mu$G would yield 
$B/B_{sc}>1$ for all bursts.

The upper histogram of Fig. \ref{fig1}, which shows the lower
limit of $\epsilon_B$ ($B/B_{sc}$) for our sample of 38 bursts, shows one peak at $\epsilon_B \approx 10^{-7} -
10^{-6}$ ($B/B_{sc} \approx 3 - 10$), where 10/38 of the bursts reside. The mean and median values of the 
histogram are $\log(\epsilon_B) \sim -5$ ($B/B_{sc} \sim 30$). The minimum and maximum values of 
the lower limit of $\epsilon_B$ ($B/B_{sc}$) are $\sim 2\times10^{-10}$ and $\sim8\times10^{-4}$
($\sim 0.1$ and $\sim 300$). 
The bottom histogram on Fig. \ref{fig1}, which shows the measurements of
$\epsilon_B$ ($B/B_{sc}$) for our subsample of 23 bursts, shows two peaks.  
One is at $\epsilon_B \approx 10^{-6} - 10^{-5}$ ($B/B_{sc} \approx 10 - 30$) with 7/23 bursts, 
and the other is at $\epsilon_B \approx 10^{-4} - 10^{-3}$ ($B/B_{sc} \approx 100 - 300$) with 5/23 bursts. 
The mean and median values of the histogram are also $\log(\epsilon_B) \sim -5$ ($B/B_{sc} \sim 30$). The
minimum and maximum values of $\epsilon_B$ ($B/B_{sc}$) are $\sim 10^{-8}$ and $\sim 0.03$ 
($\sim 1$ and $\sim 1700$).

Although we have decided to show the values of $\epsilon_B$ for a
constant density medium case in Tables \ref{table1} and \ref{table2} and 
in Fig. \ref{fig1}, we remind the reader that 
the more relevant quantity to focus on is $B/B_{sc}$ (also presented in
these tables and figure), since this quantity is density-independent. 
Let us show this explicitly now. Suppose that the GRB goes off in
a wind medium, then eq. (\ref{epsilon_B_peak_flux}) should be replaced with
the relevant equation for the wind medium, which is $f_p \propto
\epsilon_B^{1/2} A_{*} E^{1/2} t^{-1/2}$, where $A_{*}$ is the wind density
parameter (e.g., Granot \& Sari 2002). It is easy to show that 
$B/B_{sc}=(\epsilon_B/\epsilon_{B,sc})^{1/2}$ is density-independent, 
because the wind density is $n \propto E^{-1} A_{*}^2 t^{-1}$.  Therefore,
it is convenient to focus on the density-independent quantity$B/B_{sc}$.

\subsection{Consistency check of $\nu_i$}

We can also check that the measurement of $\epsilon_B$ for our subsample of 23 bursts 
is also consistent with the fact that at the time of the peak of the radio
light curve the observing radio band, $\nu_R$, should be approximately equal to the 
injection frequency. The injection frequency is given by (Granot \& Sari 2002)

\begin{equation} \label{nu_i}
\nu_i = (3.73 \times 10^{15} {\rm Hz}) (p-0.67) (1+z)^{\frac{1}{2}} E_{52}^{\frac{1}{2}} \bar{\epsilon_e}^2  \epsilon_B^{\frac{1}{2}} t_d^{-\frac{3}{2}}, 
\end{equation}
where $\bar{\epsilon_e} = \epsilon_e \left(\frac{p-2}{p-1}\right)$ and
$\epsilon_e$ is the fraction of the total energy behind the shock in
electrons, and $t_d$ is the observed time after the burst in days. We 
note that the injection frequency depends strongly on $\epsilon_e$ and $p$.
This is the main reason why we decided to use the constraint on the radio
light curve peak flux, instead of a constraint on $\nu_i$, 
to determine a lower limit on $\epsilon_B$. 

The value of $\nu_i$ in eq. (\ref{nu_i}) is valid {\it before} the jet break,
$t_j$.  After $t_j$, $\nu_i$ decrease as $\nu_i \propto t^{-2}$ independent of density medium
(Sari et al. 1999), therefore, for the case when $t_j
< t_p^{obs}$, we will calculate $\nu_i$ at $t_j$, $\nu_i(t_j)$, with
eq. (\ref{nu_i}), and its value at $t_p^{obs}$ will be given by
$\nu_i(t_p^{obs})=\nu_i(t_j)(t_p^{obs}/t_j)^{-2}$. This same behavior of
$\nu_i$ is found in numerical simulations of De Colle et al. (2012) for a
wind medium. However, for a constant density medium, $\nu_i$ steepens more
than $\propto t^{-2}$ after the jet break (van Eerten \& MacFadyen 2013), to
$\nu_i \propto t^{-2.9}$ according to De Colle et al. (2012).  Nevertheless,
the conclusions in the next paragraph apply even for this steeper behavior.

As a rough consistency check and keeping in mind the
uncertainties in calculating $\nu_i$, we check whether $\nu_i(t_p^{obs}) \sim
\nu_R$. We use the same assumptions
as before, $p = 2.4$, $E = 5 E_{\gamma}^{iso}$, and use 
$\epsilon_e \approx 0.2$ (as found in a literature compilation of $\epsilon_e$
in Santana et al. 2014).  We use the values of $\epsilon_B$ found in the previous section (Table
\ref{table2}) to find $\nu_i(t_p^{obs})$ and find that 
for 30 per cent of our sample of 23 GRBs, $\nu_R \lae \nu_i(t_p^{obs})$ ($\nu_R  = 8.5$
GHz, otherwise noted in the caption of Table \ref{table1}).  Owning to the 
strong dependence of $\nu_i$ on $\epsilon_e$ and $p$, we allow $\epsilon_e$ to
vary only by a factor of $\sim 2$, and $p$ to vary between $p=2.2-2.4$, and
find that all GRBs are consistent with $\nu_R \sim \nu_i(t_p^{obs})$ (except
one, GRB 010921, which is consistent within a factor of few). A similar conclusion can 
be found for the wind density medium case. We, therefore, 
conclude that our results on $\epsilon_B$ obtained using
the constraint on the radio peak flux are in agreement with the location of the 
injection frequency at the time of the peak, keeping in mind the uncertainties
in this calculation.

\section{Discussion and Conclusions} \label{final}

Assuming that the radio GRB light curve originates in the 
external forward shock and that its peak at a few to tens of days 
is due to the passage of the injection frequency through the radio 
band (before or after the jet break), we have found a lower limit/measurement for the 
fraction of the energy in the magnetic field to the total energy in 
the shocked fluid behind the shock, $\epsilon_B$.  This lower
limit (or measurement, for those bursts with estimates of the jet break time)
 depends on the isotropic kinetic energy in the external shock,
which is calculated assuming that the efficiency in producing gamma-rays 
is about 20 per cent.  If the efficiency is smaller (larger), then
$\epsilon_B$ would be smaller (larger) than the obtained value. 
We have also assumed that the CSM density is 1 cm$^{-3}$.  This means that 
$\epsilon_B$ in our histograms and tables can be viewed as displaying a value of the quantity
$\epsilon_B n$. We have also expressed our results as a function of the ratio of
the magnetic field behind the shock, $B$, to the expected shock-compressed CSM
field, $B_{sc}$, and we have taken the unshocked CSM field to be $B_0 \sim 10 \mu$G.  
This ratio is an important quantity since it is independent of the CSM density, 
and it indicates the level, above simple shock
compression, that the magnetic field should be amplified.  However,
$B/B_{sc}$ does depend on the assumed value of $B_0$ as $\propto B_0^{-1}$.

A clear prediction of our model is that the radio spectral index $\beta$, 
$f_{\nu} \propto \nu^{\beta}$, where $f_{\nu}$ is the specific flux, 
should be positive ($\beta=1/3$) before the radio light curve peak and then it
should become negative ($\beta = - (p-1)/2 = -0.7$ for $p=2.4$) after it, 
if the peak in the light curve is produced by the passage of the injection
frequency through the radio band. 
This can be tested for radio afterglows that have data at different radio wavelengths
before and after the light curve peak.

A possible uncertainty in our study is the fact that the peak in the radio
afterglow light curve could also have a contribution of an external reverse
shock component. Chandra \& Frail (2012), in their table 4, reported:
1. the second peak of afterglow light curves that showed two peaks and, 
2. they did not report peaks that occurred earlier than 3 days of afterglow
light curves that showed a single peak to avoid a possible reverse shock
contribution to the observed peak flux.  Nevertheless, if the peak flux in the
radio afterglow light curve that we have used has a non-negligible
contribution from the reverse shock, then the true value of  
$\epsilon_B$ in the forward shock would actually be smaller than the one we
calculated, see eq. (\ref{epsilon_B_peak_flux}). 

Another uncertainty is the fact that the jet break could have been
misinterpreted. For this reason, in addition to determining a measurement on 
$\epsilon_B$ (and $B/B_{sc}$), we have chosen also to determine a lower limit, 
for which the time of the jet break is not used.  This is a
conservative approach, given the fact that jet breaks might be more difficult
to identify as thought before (e.g., Leventis et al. 2013 and references
therein).  

Keeping in mind the uncertainties discussed in the previous paragraphs, 
we would like to determine whether the magnetic field behind the external 
forward shock in GRBs could be produced by simple shock-compressed CSM magnetic field
(as found for a small sample of {\it Fermi} satellite GRBs, see, 
Kumar \& Barniol Duran 2009, 2010; Barniol Duran \& Kumar 2011a, 2011b)
or if an additional amplification mechanism is needed. The first thing to determine 
is the value of the unshocked CSM magnetic field, which is the value of the field in the vicinity
of the GRB explosion and we have taken it to be $B_0 \sim 10$ $\mu$G.
The reason for our choice of $B_0$ is that in the Milky Way
the field is about $6$ $\mu$G near the Sun and 
several $100$ $\mu$G in filaments near the Galactic Center (Beck 2009).
Values of $5$ $\mu$G have been measured in radio-faint galaxies
(with star formation rate of SFR $\sim0.2M_{sun}$/yr), $\sim 10$ $\mu$G in normal spiral galaxies, 
$20-30$ $\mu$G in gas-rich spiral galaxies with high star formation rates
(SFR $\sim3M_{sun}$/yr), and $50-100$ $\mu$G in starburst galaxies (SFR $\sim10M_{sun}$/yr)
(Beck 2012).  The measurements of the global field of these galaxies
suggest that $B_0$ could be $B_0\lae 100\mu$G.  This means that values of up to 
$B/B_{sc} \sim 10$ could be consistent with simple shock-compression of a seed
field. This suggests that for our sample of lower limits, for 22/38 bursts (58
per cent) the shock compression origin of the magnetic field is not ruled out and remains a
viable possibility.  The same applies for 7/23 burst (30 per cent) in our subsample of
measurements.  The remaining bursts in both samples, which have larger values
of $B/B_{sc}$, would need an amplification mechanism beyond shock-compression,
unless $B_0$ is larger in these bursts for some reason.


\begin{figure*}
\begin{center}
\includegraphics[width=17cm, angle = 0, clip=true, viewport=.0in .0in 8in 3in]{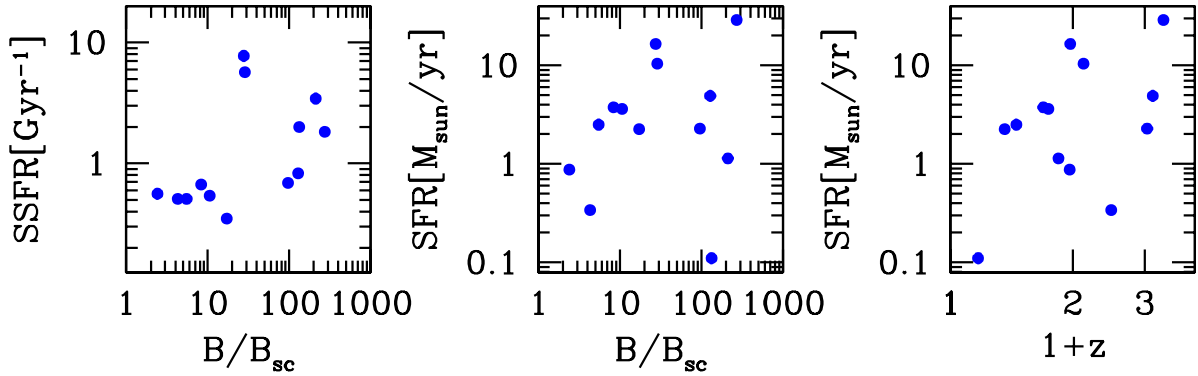}
\end{center}
\caption{GRB host specific star formation rate (SSFR) and star
formation rate (SFR) as a function of $B/B_{sc}$ (left and middle panels)
and star formation rate as a function of $1+z$ (right panel).}
\label{fig2} 
\end{figure*}


As suggested by the measurements presented in the previous paragraph (see, e.g., Beck 2009) and 
also by larger fields of $0.5-18$ mG obtained in starburst 
galaxies with SFR $\sim100M_{sun}$/yr (Robishaw, Quataert \& Heiles 2008),
galaxies with higher SFR tend to have higher global fields.  The strength of the
global field might be correlated with the field in the vicinity of the GRB, $B_0$.  
If this is the case, then simple shock-compression would remain a viable
possibility for those bursts for which we have found large values of 
$B/B_{sc}$, if they reside within galaxies with high SFR.  This idea is in the
spirit of the work of Thompson, Quataert \& Murray (2009), where they find
that for supernova remnants (SNRs) fields larger than that due to shock-compression
alone might not be needed for SNR in starbursts (nor for the average SNR in
most normal spirals, although it might be needed for individual SNRs at
particular moments of time).  

We find that 13 GRBs out of our sample of 23 have their host galaxy SFR 
reported  in Savaglio, Glazebrook \& Le Borgne (2009\footnote{The GHostS 
database can be found at www.grbhosts.org}).  We present 
these values in Table \ref{table2} and plot them 
as a function of our values of $B/B_{sc}$ and as a function of 
redshift in Figure \ref{fig2}. There is no correlation between SFR and
$B/B_{sc}$, nor SSFR (specific SFR) and $B/B_{sc}$.  This does not support the idea that
bursts with high values of $B/B_{sc}$ reside within galaxies with high SFR,
and therefore they do require an amplification beyond shock-compression.  This
assumed however that the global galactic field is a proxy of the field in the vicinity of the
GRB progenitor, $B_0$, which might not be the case with a progenitor that regulates
the medium and the magnetic fields in its vicinity via winds or other
mechanisms, or if the pre-shocked magnetic field is strongly amplified  
by instabilities that operate ahead of the shock (e.g., 
Milosavljevi\'c \& Nakar 2006).

We also report the morphology of GRB host galaxies (Wainwright et al. 2007) 
in Table \ref{table2}, using the notation in Savaglio et al. (2009).
Only 10 GRBs of our sample of 23 have reported morphologies.  Table \ref{table2}
is presented in descending order of $B/B_{sc}$.  There is no clear indication that
GRBs hosted by galaxies of different morphologies have different strengths of
$B/B_{sc}$.  However, as mentioned above, global host galaxy properties might
not represent the properties in the vicinity of the GRB.
We emphasize that the ideas discussed
in this paragraph and in the previous one are based on small number statistics
and depend on the jet break time estimate.  Larger samples in the future 
could be used to test them\footnote{There is
also the possibility of an observational bias in the radio sample; however, 
we find no correlation between $B/B_{sc}$ and redshift for the entire sample, nor for the subsample 
with SFR.}.  
 
Using X-ray and optical data for a sample of GRB afterglows, 
Santana et al. (2014) have also constrained $\epsilon_B$: an upper limit with
their X-ray sample and a measurement with their optical sample.
We find that 4 GRBs: GRB 071003, GRB 080603A, GRB 090313 (in
their optical sample), and GRB 100814A (in their X-ray sample), are common to
our radio sample. Using the same assumptions, we find that our lower limits on 
$\epsilon_B$ for these 4 GRBs are consistent with the 3 measurements of
$\epsilon_B$ (optical sample) and the upper limit on $\epsilon_B$ (X-ray
sample) in Santana et al. (2014), showing that their methods 
and ours are consistent with each other. We note that our lower limits are
only smaller by a factor of $\lae 3$ than the measurements/upper limits of
these 4 GRBs.  This means that for these 4 bursts the lower limits on
$\epsilon_B$ are a good estimate of their actual values and, thus, require
that $t_j \sim t_p^{obs}$.  Only one of these 4 GRBs is in our subsample of
GRBs with know jet break, GRB 090313.  We find that our $\epsilon_B$ estimate
using radio data is a factor of 100 larger than the one reported in Santana et
al. (2014), since for this burst $t_j \ll t_p^{obs}$. This GRB has the highest 
$\epsilon_B$ value in Table \ref{table2} (and abnormally high compared with
the rest of the sample). After inspecting its optical/X-ray afterglow, we find
that its jet break time ($<1$ day) is based only on X-ray observations, since
at this time the optical afterglow displays flares, and also contamination from a nearby
source (Berger 2009, Mao et al. 2009, Melandri et al. 2010). Therefore, this jet break can be regarded as
questionable, but we decided to keep it in Table \ref{table2} and
Fig. \ref{fig1}. 

Santana et al. (2014) find a weak correlation
between $E$ and $\epsilon_B$. We do not find it in our radio sample with
estimated jet break times, which might be due to the uncertainties in $t_j$.
Finally, the mean/median value found for $B/B_{sc} \sim 30$ in our radio
sample (for both the lower limits and the measurements) agrees very well with
the value found in Santana et al. (2014), where they report $B/B_{sc} \sim
50$. We note that three very different methods (using radio, optical and X-ray data) yield 
consistent values of the level of amplification needed in GRB relativistic 
collisionless shocks.

Lemoine et al. 2013 have studied bursts with $>100$ MeV (detected by the 
{\it Fermi} satellite), X-ray, optical and radio data. They explain all these
observations in the context of the external forward shock (see, e.g., Kumar \&
Barniol Duran 2009). However, they consider a decaying field behind the shock
(Lemoine 2013), where the $>100$ MeV emission is produced by electrons close
to the shock front, where $\epsilon_B \sim 10^{-2} - 10^{-1}$ and the X-ray,
optical, and radio emission is produced by electrons further downstream of the
shock front, where $\epsilon_B \sim 10^{-6} - 10^{-4}$. The latter value of
$\epsilon_B$ is consistent with our obtained median of $\epsilon_B \sim
10^{-5}$ using radio data.

We note that the median value of $\epsilon_B$ found in this work, 
by assuming $n=1$ cm$^{-3}$ and a 20 per cent efficiency in producing 
the prompt emission gamma-rays, is a factor of $\sim 600$ smaller than 
the median value found in, e.g., Panaitescu \& Kumar (2002), in which a detailed multiwavelength 
analysis of several of the bursts in our sample are presented. The main difference 
is that the median density found in Panaitescu \& Kumar (2002) is $\sim 10$ times larger 
and the median efficiency is $\sim 70$ per cent. Also, the numerical pre-factor 
used in eq. (\ref{peak_flux}) in Granot \& Sari (2002) is much larger than the one used 
in Panaitescu \& Kumar (2002), and since 
eqs. (\ref{epsilon_B_peak_flux}) and (\ref{epsilon_B_peak_flux_2}) depend on this 
pre-factor squared, then the effect is to lower $\epsilon_B$ considerably. A similar
conclusion applies to our results regarding $B/B_{sc}$.

We would like to finish by comparing our values of $B/B_{sc}$ with the ones
found for young shell-type supernova remnants (SNRs). There is some evidence
for field amplification in Galactic SNRs with levels of $B/B_{sc} \sim 10-100$ 
at particular moments of time (e.g., V\"olk et al. 2002, 
V\"olk, Berezhko \& Ksenofontov 2005, Berezhko, 
Ksenofontov \& V\"olk 2006, see, also, recently, Castro et al. 2013).  It is 
interesting to note that a similar level of amplification as the one presented 
in this work and in Santana et al. (2014) is observed in these systems. Taken
at face value, this might indicate a common amplification mechanism in GRB
relativistic afterglows and non-relativistic SNRs, which might even be
found in other very different systems.  We should caution, however, that the
environment in the vicinity of the GRB explosion remains uncertain and every
possible effort to characterize it should be made.

This work emphasizes the importance of radio afterglow observations in
studying relativistic collisionless shocks. By using the simple method
presented here, one can extract valuable information about the magnetic field
behind the shock for GRBs, which have not been monitored extensively at other 
wavelengths to allow for a complete and detailed afterglow modeling. Radio
monitoring of GRB afterglows at various wavelengths continues to be an
indispensable tool in the study of collisionless shocks.

\section*{Acknowledgments}

I dedicate this work to the memory of Reinaldo Ca\~nizares Pesantes.
It is a pleasure to thank Jessa Barniol for her help with Table 1.
I thank Rodolfo Santana and Pawan Kumar for a careful reading of the manuscript, 
and also thank Paz Beniamini, Yuval Birnboim, Fabio de Colle, Uri Keshet, Ehud
Nakar, Tsvi Piran, Rongfeng Shen and Hendrik van Eerten for useful discussions.
This work was supported by an ERC advanced grant (GRB) and by the I-CORE Program
of the PBC and the ISF (grant 1829/12). This research has made use of the 
GHostS database (www.grbhosts.org), which is partly funded by Spitzer/NASA 
grant RSA Agreement No. 1287913.

\end{document}